\documentstyle[epsf]{l-aa} 
\begin{document}

\newcommand{\mic}{\,{\rm \mu m} } 

\title{FIRBACK: I. A deep survey at 175 $\mic$ with ISO, preliminary results} 

\author{J.L. Puget  
          \inst{1} \and 
          G. Lagache 
          \inst{1} \and 
          D.L Clements
          \inst{1} \and
          W.T. Reach
          \inst{2} \and
          H. Aussel
          \inst{3} \and
          F.R Bouchet
         \inst{4} \and
          C. Cesarsky
          \inst{3} \and 
          F.X. D\'esert
          \inst{5} \and	
	  H. Dole
	  \inst{1} \and
          D. Elbaz
          \inst{3} \and 
          A. Franceschini
          \inst{6} \and
          B. Guiderdoni
          \inst{4} \and
          A.F.M Moorwood
          \inst{7} } 

\institute{$^1$ Institut d'Astrophysique Spatiale, B\^at.  121, 
Universit\'e Paris XI, 91405 Orsay Cedex, France\\
$^2$ IPAC, California Institute of Technology, Pasadena,
California 91125, US \\
$^3$ Service d'Astrophysique/DAPNIA/DSM, CEA Saclay, 91191 Gif sur Yvette,
France\\
$^4$ Institut d'Astrophysique de Paris, 98 bis Boulevard ARAGO,
75014 Paris, France\\
$^5$ Observatoire de Grenoble, BP 53, 
414 rue de la piscine, 38041 Grenoble Cedex 9, France  \\
$^6$ Osservatorio Astronomico di Padova, 35122 Padova, Italy\\
$^7$ ESO, Karl-Schwarzschild-Str, 2, D-85748 Garching bei Munchen}

\date{Received 15 July 1998; Accepted }  
          
\maketitle

\section{Abstract}

FIRBACK is a deep survey conducted with the ISOPHOT instrument aboard the Infrared
Space Observatory (ISO) at an effective wavelength of 175 $\mu$m. 
We present here results we have obtained on the first field, the so-called
Marano1 which covers around 0.25 square degree. We 
find that the source 
density for objects with a flux above 200 mJy exceeds 
the counts expected for sources found in the IRAS deep surveys with 
a similar flux by about an order of magnitude.
Such an excess was expected on the basis of the high far infrared 
background detected with the FIRAS and DIRBE instruments aboard the Cosmic 
Background Explorer (COBE).
These sources are likely to be redshifted infrared galaxies.
The steep number counts indicate strong cosmological evolution in this population.
The detected sources account for only 10 $\%$ of the cosmic IR background.
An extrapolation of the counts down to about 10 mJy would be needeed to account for
the whole background
at this wavelength.

\section{Introduction}

New windows are opening in the nearly unexplored evolution of
distant galaxies at infrared and submillimetre wavelengths.
Significant progress in the knowledge of IR/submm emission of moderate --
and high -- redshift sources is now possible. Deep surveys in the mid infrared
with the Infared Space Observatory camera (ISOCAM) (Cesarsky et al. 1996)
and in the far infrared with the photometer ISOPHOT (Lemke et al. 1996) 
were designed to
yield strong constraints on galaxy evolution at these wavelengths.
In this paper we present the first results of a deep survey with ISOPHOT
at $\lambda_{eff}=175 \mu$m. 

For some time, astronomers have suspected the importance of 
the ``optically dark'' side of galaxy evolution where 
stellar UV/optical light due to star formation is reprocessed to thermal
radiation by dust
(for example Setti and Woltjer 1970;  Stecker, Puget and Fazio 1977).
The IRAS all-sky survey unveiled a 
local population of infrared galaxies ranging from normal spirals to the 
spectacular luminous and ultraluminous infrared galaxies 
(resp. LIRGs and  ULIRGs), but only a handful of these sources had
a significant redshift. Thus,
our knowledge of the dark side of galaxy evolution has so--far been
limited to local or 
low--redshift galaxies. At present about one third of the bolometric 
luminosity released by stars finally escapes galaxies at IR/submm 
wavelengths (Soifer and Neugebauer, 1991).
A large fraction of this is associated with star formation.
With their limited sensitivity, deep IRAS
surveys were unable to probe redshifts 
larger than $\sim 0.3$ (Ashby et al. 1996), although they already show a 
significant evolution in number counts (see e.g. Bertin et al. 1997).

A host of observational evidence seems to directly demonstrate
the presence of 
dust at higher z. For instance, IRAS identified 
IRAS F10214+4724,
a very peculiar, ``hyperluminous'' galaxy at $z=2.286$
(Rowan--Robinson et al., 1991). This object was
detected by IRAS because it is strongly magnified by gravitational lensing 
(Serjeant et al. 1995, Eisenhardt et al., 1996).
It probably has many counterparts which
are so far undetected. About twenty radiogalaxies 
and quasars have also been observed at submm/mm wavelengths 
(see e.g. Hughes et al.,
1997; and references therein). Although their evolutionary status is not 
completely understood, the most likely interpretation is that the submm 
emission is mostly due to dust heated by star formation in the host galaxies.

A spectacular breakthrough was recently achieved with the detection of the
long--sought ``Cosmic Infrared Background'' (hereafter CIRB) in
FIRAS residuals between 200 $\mu$m and 2 mm 
(Puget et al., 1996; Guiderdoni et al., 1997, hereafter GBPLH; 
Fixsen et al., 1998; Lagache et al., 1998), and in DIRBE residuals at
140 and 240 $\mu$m (Schlegel et al., 1997; 
Hauser et al., 1998; Dwek et al., 1998; Lagache et al., 1998). 
The background is produced by the line--of--sight accumulation of
extragalactic sources. The level of the 
CIRB is significantly higher than the lower limits on the
``Cosmic Optical Background'' (hereafter COB) given by summing
faint galaxy counts that probably lie close to convergence (Williams et al., 1996).
This is a strong evidence that a significant fraction of star formation 
is hidden from optical observations and observable only at far infrared and 
submillimeter wavelengths.

With the rapid development of the case for dust at high redshift,
theoretical efforts have led to detailed modelling of spectral
evolution at IR/submm wavelengths (Franceschini et al. 1991, 1994,
1998). These spectral energy distributions can be consistently implemented
in the general framework of
hierarchical clustering (GBPLH; Guiderdoni et al., 1998). The generic
prediction of these models is a high level of IR/submm emission due to the 
continuous formation of large galaxies through the merging of smaller ones. 
This model has allowed GBPLH to predict the decomposition of the 
CIRB into faint galaxy counts. A family of scenarios was generated that
agree with local IR constraints (IRAS luminosty functions and faint 
counts), and various levels of evolution consistent with the 
observed CIRB in FIRAS residuals.
The authors concluded their analysis by showing that, counts
at 15 $\mu$m and 60 $\mu$m were relatively insensitive to the details of the 
scenarios, but that submm counts could provide constraints strong enough to
break the degeneracy. There are three reasons for this specific behaviour: 
(1) The rest-frame wavelength range around 100 $\mu$m encompasses the bulk
of IR emission due to star formation, while the 15 $\mu$m band probes
the fraction of radiation absorbed by the Polycyclic Aromatic Hydrocarbons 
which is less directly related to the total output and thus not a reliable
tracer of star 
formation. (2) The cosmic star formation rate density directly seen at UV 
wavelengths strongly increases between $z=0$ and 1.5 (Lilly {\it et al.} 1996). 
It is likely that similar 
behaviour appears in the IR. (3) Because of the shape of the spectral 
energy distribution, the 
$k$--correction is {\it negative} at 175 $\mu$m and the detection of galaxies
at $z \sim 1-2$ is favoured (Blain \& Longair, 1996).

Consequently, GBPLH proposed the concept of a deep ISOPHOT
survey with the C160 filter (effective wavelength 175 $\mu$m).
To provide sufficient constraints on the models, the exposure times
per pointing had to reach $\sigma_{noise} \sim 10$ to 20 mJy.
This deep ISOPHOT survey was expected to yield a probe of galaxy evolution 
complementary to deep ISOCAM surveys. In the IR range, the ISO deep surveys
follow-up has given number counts at 15 $\mic$
(Oliver et al., 1997; D\'esert et al., 1998; Aussel et al., 1998)
and 7 $\mic$ (Taniguchi et al., 1998).
Kawara et al. (1998) have conducted a deep survey at 175 $\mic$ in the
Lockman Hole region.
At the other end of the submillimetre decade, the first deep 
survey obtained with SCUBA on the JCMT has also unveiled a 
population of submm sources with number counts much larger than 
the no--evolution predictions (Smail et al., 1997,1998; Hughes et al., 1998;
Barger et al., 1998). 

Herein we present the first results of this ISOPHOT survey as implemented
in the 
Marano field. An area of 0.25 square degree was covered and a detection level of 
$\sim$100 mJy (5$\sigma$) was reached. As anticipated from the level of the CIRB,
a large number 
of sources was found, much larger than the predictions based on IRAS 
luminosity functions without strong evolution. Section 3 summarizes our observations.
Section 4 explains the data reduction and calibration methods. Section
5 gives the resulting maps and source catalogs. These results are discussed 
in section 6, which compares the observed number counts with model predictions 
and IRAS 60 $\mic$ counts. Section 7 provides our 
provisional conclusions. Forthcoming papers in this series will present 
the complete survey, elaborate the constraints on the models and detail the 
identification and follow-up strategy and results.

\section{Observations}

The observations were performed at an effective wavelength of 
175$\mic$ with the C200 array (2x2 pixels) of ISOPHOT. 
The pixel field of view is 1.5 arcmin.
The AOT PHT22 was used with
the C160 filter. The observations consist of four 19x19 rasters centred on
(RA,DEC) in J2000 equal to (03h13m25.6s, -55d03m43.9s), (03h13m9.6s, -55d01m26.0s), 
(03h12m53.6s, -55d03m43.9s) and (03h13m9.6s, -55d06m25.4s) respectively.
Each raster was performed in the spacecraft (Y,Z) coordinate system which
is parallel to the edges of the detector array. The field area covered by each
raster is about 30'x30'. The exposure time is 16s per pixel.
Rasters were performed with one pixel overlap in
both Y and Z direction. Therefore, for the maximum redundancy region, the
effective exposure time is 256 sec per sky position (16 sec per pixel,
4 rasters and redundancy of 4).
Displacements between the four raster centers correspond to 2 pixels. This mode of
observation, which does not provide proper sampling of the point-spread-function (PSF),
was chosen to assess
the reliability of faint source observations with ISOPHOT at a time when 
this was in question.

\section{Data reduction}

We use PIA, the ISOPHOT interactive analysis software 
version 6.4 (Gabriel et
al., 1997), to correct for instrumental effects, the glitches induced
by cosmic particles and to provide an initial calibration. First we apply
the non-linearity correction due to 2 independent effects: the 
non-linearity of the Cold Readout Electronics and the downward curving
ramp due to de-biasing. Deglitching is performed for each individual
ramp and then the mean signal per position is derived by averaging
the linear fit on each ramp (see Gabriel et al., 1997 for details).
The signal shows a long term drift
which is corrected by using the two internal calibration
(FCS) measurements bracketing each individual observation. 
Inside each raster, the long term drift represents less than 10$\%$ 
of the signal. The calibration is performed by deriving the mean value of 
the two FCS measurements. The contribution of the long term drift 
is thus lower than 5$\%$. Flat fielding is achieved using the high 
redundancy in the measurements: we compare the fluxes over all positions
(which have a redudancy of 4)
and derive the difference in responsivity between the four pixels
for each raster.
Glitches inducing long term drifts are also corrected.
The flux is
finally projected on a 5"x5" coordinate grid using our
own projection procedure.

We compare the average surface brightness as given by PIAv6.4  
with that derived from 
an interpolation of DIRBE data at 100, 140 and
240$\mic$. The value found by PIA is a factor 2.5 greater than that from 
DIRBE.
This discrepancy results mainly from the fact that the solid angle used for the
extended source 
calibration in PIAv6.4 is the solid angle of one pixel and not the
integral of the Point Spread Function (PSF). 
When the solid angle is computed using the PSF wing measurements
on Saturn (Toth et al., 1997), it is equal to 2.2 times the solid angle of the pixel. 
The PSF determined in this way is shown Fig. 1. 
When we correct the PIA calibration by this factor, this 
leads to a calibration difference with DIRBE of only $12\%$,
well within the combined uncertainties of the PHOT point source calibration 
and the solid angle. We conclude that
the PHOT calibration established with strong sources can 
be extrapolated to lower
fluxes, such as the flux from the background, 
but that the proper PSF must be taken into account.
In the present work, we thus adopt a calibration based on the DIRBE fluxes.
Errors estimates given by the PIA are much smaller than the systematic errors.
The two mains ones are the photometric calibration uncertainty itself, which cannot be larger
than 15$\%$ on the basis of the good agreement with the DIRBE brightness, and
the uncertainty on the PSF. The PSF we use is based on the Saturn measurements and differs from
the models (used in the most recent version of the PIA) by a factor around 1.3. Alltogether, 
for point sources, we thus estimate an uncertainty of about 30 $\%$.

\begin{figure}  
\epsfxsize=8.cm
\epsfysize=7.cm
\hspace{2.cm}
\vspace{0.0cm}
\epsfbox{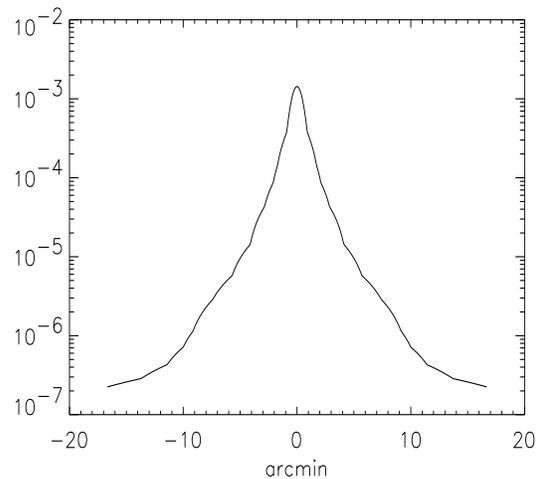}
\caption{ISOPHOT Point Spread Function at 175 $\mic$ (C-160) combining
Saturn measurements in the wings with measurements of the central part of
the main beam (Toth et al., 1997)} 
\end{figure}

\section{Results}

Fig. 2 shows 4 images each made from one individual raster.
The comparison of these images shows a high degree of correlation.
The same large scale structure is seen in the four rasters. This
structure corresponds to the edge of a cirrus cloud already
detected with IRAS data at 100 micron (Fig. 3). 
Smaller structures are also very well reproduced in the four 
rasters. These small scale structures are due to the 
combination of point sources, the background of unresolved sources
and cirrus fluctuations. 
Differences between independent raster maps yield a residual 
rms fluctuation of 0.055 MJy/sr. This leads to a noise in the final
four coadded maps 2$\sqrt{2}$ times smaller or 1.96 10$^{-2}$ MJy/sr 
(3.7 mJy per 92" pixel) which is representative
of the detector noise, but only an upper limit since it includes 
some residual differences in the response between individual rasters.
The brightness variations in the maps have
a rms value of 0.13 MJy/sr and are probably dominated by the cirrus component.
Cirrus clouds are known to have a steep power spectrum (Gautier et al., 1992) 
proportional to $k^{-3}$ and thus gives strong brightness variations only at 
large scales.  
The brightness is on average equal to 3.16 MJy/sr and varies from 2.9
to 3.3 MJy/sr for the diffuse emission. 
This brightness is composed of an extragalactic background of 1 MJy/sr
(Lagache et al., 1998), zodiacal emission around 0.7 MJy/sr (based on the W.T. Reach model,
private communication)
and cirrus of around 1.46 MJy/sr. Most of the large scale variations is attributed to the cirrus
component which varies from 1.2 and 1.6 MJy/sr.
Point-like sources are also clearly visible.

\begin{figure*}  
\epsfxsize=18.cm
\epsfysize=25.cm
\hspace{0.0cm}
\vspace{0.0cm}
\epsfbox{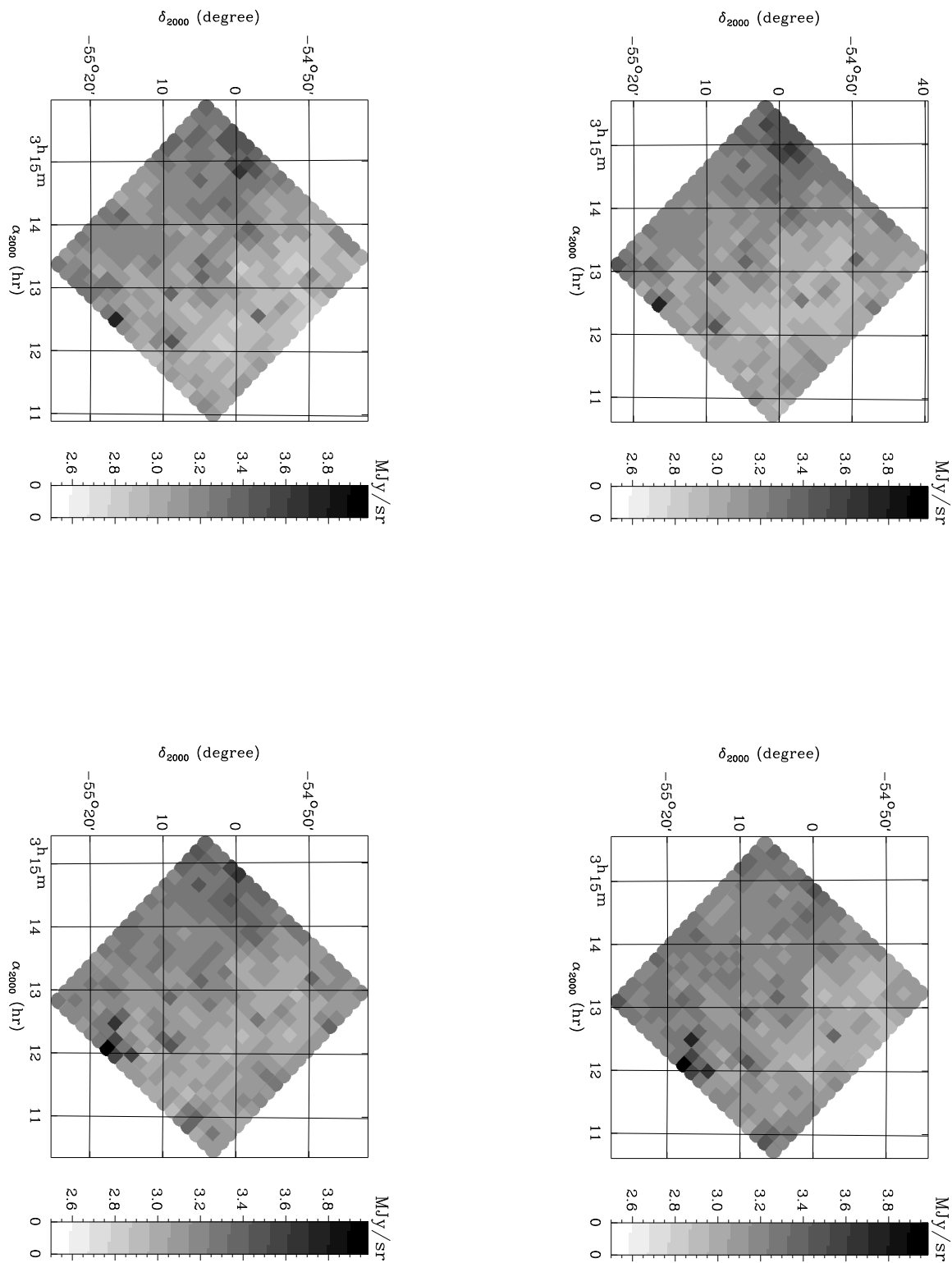}
\caption{Images of the four rasters. Note the displacement of two pixels
between raster centers.} 
\end{figure*}

\begin{figure*}  
\epsfxsize=22.cm
\epsfysize=16.cm
\hspace{-4.2cm}
\vspace{-2.0cm}
\epsfbox{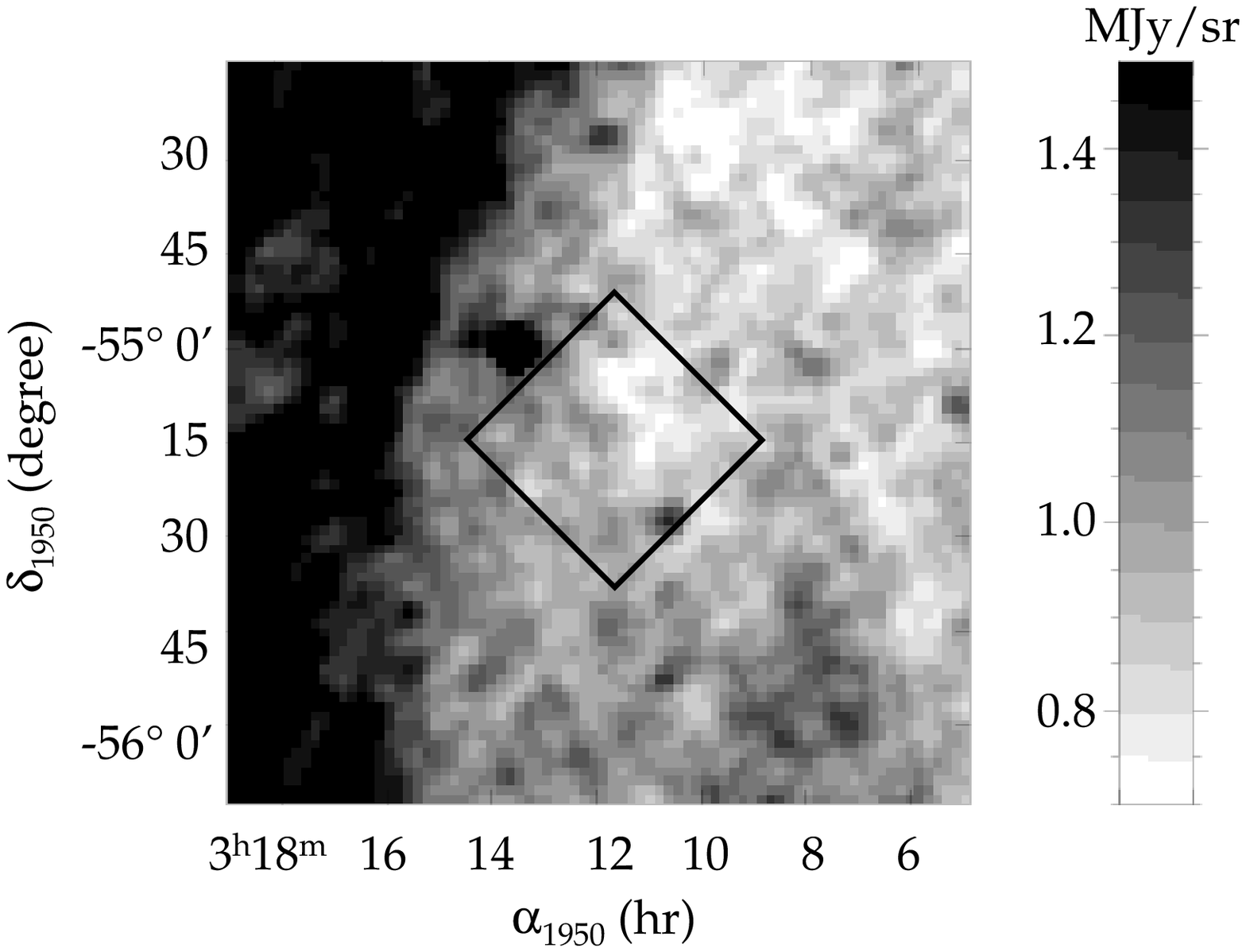}
\caption{Marano field observed with ISOPHOT at 175 $\mic$ projected on the
IRAS 100 $\mic$ map. We have added to the IRAS map 0.7 MJy/sr to account for
the different zero level between IRAS 100 and DIRBE 100 $\mic$.} 
\end{figure*}

We extract these sources by comparing the contrast between each pixel 
on the sky with the surrounding ones. We keep only sources 
for which the contrast is greater than 0.03 MJy/sr, the signal to noise ratio
greater than 3 and sources seen in at 
least 3 rasters in the region of maximum
redundancy (observed 4 times)
and in all rasters available for regions on the edges.
The main limitation to the photometry of these sources
is the poor PSF sampling.
The determination of the flux is done
by using only the central part of the PSF. We do aperture photometry; we derive 
the flux within an inner radius of 60" and remove a reference background 
computed between the inner radius of 60'' and an outer radius of 92". 
With the PSF such as the one displayed in Fig.1, the reference background is strongly
contaminated by the source. 
The total flux from a source is 4.75 times the measured one obtained with the procedure
described just above. 
Furthermore, the fluxes are also 
multiplied by 1.1 to correct for the transient effect estimated to be about 
$10\%$. Thus, all the measured fluxes (and therefore the noise
mesurements) must be multiplied by 4.75 and 1.1.\\

\begin{table*} \caption{\label{tab1}RA, DEC, flux and signal to noise ratio
 of the detected sources. n1 represents the 
number of rasters in which the source should 
be detected and n2 the number of rasters in which the source is
in fact detected.}
\begin{flushleft}
 \begin{tabular}{c|l|l|c|c|c|c|l} 
 & RA (J2000) & DEC (J2000) & Flux (mJy) & S/B & n1 & n2 & 
Comments     \\ \hline
0 & 3 12 5.0 & -55 17 44.4 & 849 & 37.9 & 2 & 2 & region 2. Large spiral
galaxy ESO155-10 \\
1 & 3 12 28.9 & -55 16 40.5 & 479 & 21.4 & 4 & 4 & region 2\\
2 & 3 11 59.3 & -55 14 30.3 & 375 & 16.7 & 2 & 2 & region 2\\
3 & 3 12 33.3 & -54 57 5.9 & 337 & 15.0 & 4 & 4 & isolated\\
4 & 3 12 8.5 & -55 08 55.2 & 323 & 14.4 & 4 & 4 & isolated\\
5 & 3 13 5.6 & -55 26 50.2 & 312 & 13.9 & 1 & 1 & isolated\\
6 & 3 12 53.9 & -55 09 5.3 & 265 & 11.8 & 4 & 4 & isolated\\   
7 & 3 13 11.3 & -54 49 33.8 & 246 & 11.0 & 4 & 4 & isolated\\
8 & 3 12 26.3 & -54 47 7.4 & 212 & 9.5 & 2 & 2 & isolated \\
9 & 3 14 50.0 & -54 59 33.6 & 201 & 8.9 & 4 & 4 & region 1\\
10 & 3 10 53.4 & -55 06 25.4 & 199 & 8.9 & 2 & 2 & region 3\\
11 & 3 14 8.8 & -55 15 58.3 & 191 & 8.5 & 2 & 2 & isolated \\
12 & 3 13 9.8 & -55 04 50.2 & 184 & 8.2 & 4 & 4 & isolated\\
13 & 3 15 18.9 & -55 01 37.2 & 180 & 8.0 & 2 & 2 & region 1\\
14 & 3 13 6.4 & -55 17 47.1 & 151 & 6.7 & 4 & 4 & not very isolated\\
15 & 3 14 57.1 & -54 58 11.6 & 147 & 6.6 & 2 & 2 & region 1\\
16 & 3 13 50.5 & -54 50 53.6 & 146 & 6.5 & 4 & 4 & isolated\\
17 & 3 14 57.6 & -55 00 48.4 & 142 & 6.3 & 4 & 4 & region 1\\
18 & 3 10 45.1 & -55 03 15.8 & 135 & 6.0 & 1 & 1 & isolated\\
19 & 3 12 41.1 & -54 53 49.4 & 134 & 6.0 & 4 & 4 & isolated\\
20 & 3 14 40.3 & -55 05 6.1 & 126 & 5.6 & 4 & 4 & in cirrus emission\\
21 & 3 13 55.3 & -54 58 21.8 & 124 & 5.5 & 4 & 3 & not very isolated; double radio source ID\\
22 & 3 13 25.4 & -55 04 41.6& 108 & 4.8 & 4 & 3 & Near source number 12\\ 
23 & 3 12 49.1 & -55 03 45.2 & 107 & 4.7 & 4 & 4 & isolated \\\end{tabular}\\
\end{flushleft} \end{table*}

Table 1 gives the catalog of the extracted sources. We have defined three
regions corresponding to higher IRAS 100 $\mic$ fluxes
in which several ISOPHOT sources are detected. These regions
are dominated by source confusion and not by the cirrus confusion. 
The first one (region 1 
in Table 1) is located around RA=3h14mn50s and DEC=-55d0'48'', the second
(region 2) around RA=3h12mn20s and DEC=-55d16'00'' and the third (region 3) 
around RA=3h11mn00s and DEC=-55d06'00''. 
Region 2 corresponds to the IRAS faint source
F03108-5528 and is likely associated with the large spiral galaxy
ESO155-10. In this region, 3 sources are detected with ISOPHOT. Because 
of the poor sampling, after extraction of the brightest sources (24 sources 
included in the catalog), the extraction of potential weaker sources 
becomes very dependant on the 
poorly determined positions of the bright sources (eg. sources
number 15, 17 and 22). For this reason, more sources are
probably detectable with a similar integration time and proper sampling.
Although we do not expect
a significant number of spurious sources due 
to small scale cirrus structures for a k$^{-3}$ cirrus power spectrum, we
test this conclusion by comparing the 
number of sources detected in the two halves
of the area observed defined by 
their cirrus content. We separate the map into two equal parts:
the first where the cirrus flux is greater than the median flux of 
the map, and the second where
the flux is lower. In the first part, 
we detect 13 sources and in
the second, 11 sources. We do not detect significantly 
more sources in the brightest part of the map. 
The level of cirrus clouds fluctuations is known to increase with the
average brightness to the 3/2 power (Gautier et al., 1992). 
For the variations of the cirrus brightness given above, we would expect a 50$\%$
excess in the cirrus fluctuation over the map. The excess found is only 18$\%$, which
is significantly smaller. This result is in agreement with the estimate of the cirrus
confusion noise quoted below which predicts a negligeable number of spurious 
Galactic sources.\\

The measurement noise of the sources is estimated by computing the flux of the 
sources in the 16 independent maps obtained in the maximum redundancy
region. We 
find a median rms noise of 2.48 mJy in the measured
signal through our spatial filter which leads to an rms noise 
of 12.96 mJy on the source fluxes. This
includes the detector noise but not the confusion noise 
from the combination of cirrus and weak sources.
To estimate the total noise, we measure the rms fluctuations on the map after
convolution with our aperture photometry filter.
The histogram of this flux, centred on 0, has a dispersion 
of 4.30 mJy which is a
good estimate of the total noise. Thus, the total noise in the measured 
fluxes is 22.47 mJy;
detector noise in this measurement being 12.96 mJy, 
the total sky confusion noise is thus about 18.35 mJy.
This shows that sky confusion is the dominant source of noise.
For the fraction of the brightness attributed to the cirrus
(1.46 MJy/sr on average) and assuming a power spectrum in k$^{-3}$
(Gautier et al., 1992), for the spatial filter used, we predict
a contribution to the sky confusion noise of only 4.70 mJy for the cirrus out of
a total of 18.35 mJy. A detailed analysis of the various noise components
has been performed (Lagache, 1998) and
will be presented in a forthcoming paper. For sources at the five sigma level,
we don't expect any isolated sources to be of Galactic cirrus origin.
Howevever, in region 1, where there is a peak of emission, one could question
if the sources could be due to a strong cirrus peak. In this unlikely hypothesis, 
this would reduce at most the number of sources from 24 to 21.

\section{Discussion}

\subsection{Galaxy counts}
Fig. 4 shows the number counts obtained for sources above 120 mJy ($5\sigma$ limit,
22 sources) 
and 200 mJy (10 sources). One sigma uncertainties on the fluxes 
are equal to 22.47 mJy. 
Uncertainties in the number counts are the statistical errors.

\begin{figure*}  
\epsfxsize=12.cm
\epsfysize=9.cm
\hspace{2.cm}
\vspace{1.cm}
\epsfbox{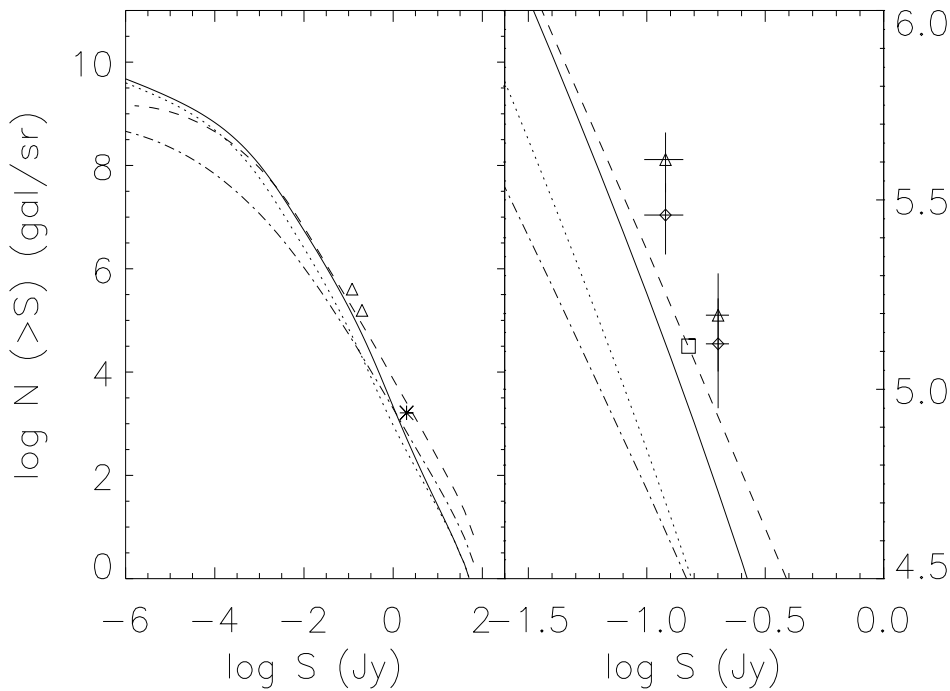}
\caption{Number counts for sources above 120 and 200 mJy ($\diamond$)
and corrected number counts ($\triangle$) together with model
predictions: Guiderdoni et al. (1997), E: with ULIRGs (continuous) and A: without
ULIRGs (dot) models, and
Franceschini et al. (1998), no evolution (dash-dot) and evolution (dash) models.
A blow up of the most relevant part of the graph is shown on the right pannel. 
We have also reported the raw number counts ($\Box$)
from Kawara et al. (1998), and from the serendipity survey ($\ast$;
Stickel et al., 1998).}
\end{figure*}

The signal to noise of our observation is large as shown by the reproductibility
(see Fig. 2) and the excellent agreement between the flux of the sources measured in the
four independent rasters (Sect.5). Thus, the main problem for the completness is
dominated by the effect of sources falling close to pixel edges, which can be
computed exactly, and the confusion between sources. The limited area in the present survey
do not allow a proper evaluation of the incompleteness due to sky confusion.
Thus, we have computed
the incompleteness correction for the beam pattern given in Fig. 1 and the
contrast criterium given in Sect.5. 
For example at 120 mJy, 41$\%$ of sources are lost, 
and 20$\%$ at 300 mJy. To correct the integral counts, we need to know
the slope which is 
not strongly constrained by our limited statistic.
 We show in section 6.2 that the IRAS counts at 
60 $\mu$m from the deep ecliptic pole survey
(Hacking \& Houck, 1987) and the colors of the IRAS galaxies 
can be used to demonstrate that counts at 175 $\mu$m must have a slope steeper
than 1.5 to converge towards the IRAS counts extrapolated to 175 $\mu$m.
Using the serendipity number counts at 175$\mic$ (Stickel et al.,1998), 
we obtain a slope of -1.9.
We choose to use a conservative 1.5 slope (that expected for a Euclidean 
distribution) to fit our counts in order to compute our 
incompleteness correction on the integral counts.  We
use the following representation for the uncorrected data:\\
 
N($>S$) = 2.89 10$^5$ $\left( \frac{S}{S_0} \right)^{-1.5}$\\

with $S_0$ = 120 mJy.

After correction for incompleteness the counts become\\

N($>S$)= 4.04 10$^5$ $\left( \frac{S}{S_0} \right)^{-1.8}$\\

valid between 120 and 200 mJy.\\

Raw number counts in the Lockman Hole region, reported in Kawara et al. (1998),
are a factor 1.6, in number, or 1.4 in flux below our uncorrected counts. 
This discrepancy is comparable with the statistical errors. Furthermore,
significant differences in calibration are likely to be present as
Kawara et al. have used the extrapolation of the spectrum of an
IRAS source to calibrate their point source fluxes.\\
Fig. 4 also shows predicted number counts from four models 
already published (Franceschini et al., 1998 and
Guiderdoni et al., 1998).
Our number counts are above the two minimal models from Franceschini et al. (1997) 
with no evolution and Guiderdoni et al. (1998) with no Ultra-Luminous IR Galaxies
(ULIRGs),
by a factor close to 10. Therefore, 
the survey shows clear evidence for strong evolution of galaxies
at 175 $\mic$. 
Our counts are above model E  of Guiderdoni et al. (1998) by a factor of 3,
and the evolution model of Franceschini et al. (1997) by a factor 2.
Sources above 120 mJy contribute for $\sim$ 0.1 MJy/sr which represents
10 $ \% $ of the best determination of the submillimeter extragalactic 
background found in the FIRAS data 
(GBPLH; Fixsen et al., 1998; Lagache et al., 1998). An extrapolation
of the counts given above finds that sources
brighter than 8.5 mJy can provide the whole background.\\ 

\begin{table}[here]
\caption{\label{tab3} Colors ($f_{\nu}$) and luminosities of FIR galaxies}
\begin{flushleft}
 \begin{tabular}{l|c|c|l} 
name & 60/175 color & Log(L$_{ir}$) (L$_{\odot}$) & Nature \\ \hline
NGC 4102$^{1}$ & 2.3 & 10.0$^{2}$ & Starburst \\
NGC 4418$^{1}$ & 1.8 & 10.7$^{2}$ & \\
NGC 6000$^{1}$ & 1.7 & 10.6$^{2}$ & Starburst \\ 
Mrk 231$^{1}$ & 2.7 & 12.1$^{2}$ & Seyfert \\ 
NGC 6090$^{3}$ & 0.7 & 10.6 & Merger \\
ARP 220$^{4}$ & 2.2 & 12.1 & Merger \\
NGC 6243$^{4}$ & 2.0 & 11.7 & Merger \\
ARP 244$^{4}$ & 0.8 & 10.8 & Merger \\ \hline
\end{tabular}\\
$^{1}$ Roche and Chandler (1993) \\
$^{2}$ Roche et al. (1991)\\
$^{3}$ Acosta-Pulido et al. (1996) \\
$^{4}$ Klaas et al. (1998) \\
\end{flushleft} \end{table}

\subsection{FIR properties}
Deep IRAS galaxy counts at 60$\mic$ go down to 110 mJy (Bertin et al., 1997).
These sources are dominated by low redshift sources (z$\le$0.2). 
Such nearby galaxies have been shown to have a far-IR Spectral Energy 
Distribution dependent on
the luminosity (Soifer and Neugebauer, 1991). 
Using a combination of IRAS, ISO and
ground based measurements, 
we deduce a 60/175 color ratio for a number of galaxies (Table 2). 
The color ratios go from 0.7 to 2.7. 
Using this constraint on the colors,
 $ {F_\nu (175) \over {F_\nu (60)}} \le 1.5 $,
appropriate for starburst galaxies, 
and galaxy counts from Lonsdale et al. (1990) and Bertin et al. (1997), 
one can make a crude prediction for the 175 $\mic$ counts 
due to low redshift IRAS galaxies. The number of galaxies
at 175 $\mic$ induced by these low redshift IRAS galaxies ($n_{175}$)
is: \\

$n_{175} (200 mJy) \le n_{60} (133 mJy) = 1.3 \times 10^{4} \quad sr^{-1} $\\

where $n_{60}$ is the number of IRAS galaxies.
At 200 mJy, this contribution from 60 $\mu$m detected IRAS galaxies 
is lower than the 175 
$\mic$ counts by a factor larger than 10. 
Measurements at the same wavelength from the Serendipity ISOPHOT
survey (Stickel et al., 1998), as shown in Fig.4, gives a mean 175 over 60 color ratio
around 2. This sample selected on the 100$\mic$ data give an upper limit 
for the color ratio. Even using this ratio, we still conclude that
the 175$\mic$ counts are a factor around 10 greater than the IRAS extrapolated
counts.

\section{Conclusion}

We have conducted a deep survey at 175 $\mu$m in search of the
objects responsible for the recently discovered CIRB. Our survey finds
a large population of point sources. Number counts for this population
are considerably in excess of those expected for the local FIR population
discovered by IRAS, but are in line with models that account for
the CIRB.
We have shown that most of the 175 $\mic$ sources detected 
in this survey have a much higher
$F_\nu (175) \over F_\nu (60)$ ratio than the IRAS galaxies. 
These sources are thus
either a new population
of nearby very cold galaxies or, more likely, are more distant sources 
(at typical redshifts 1 to 2). The models (Franceschini et al., 1998, 
and Guiderdoni et al., 1998) predict
redshift distributions with a median around 1.
In view of the steepness of the counts, the sources which account 
for the bulk of the far infrared background probably have fluxes 
around 10 mJy. Several followup programmes are underway to identify 
counterparts of these 175 $\mu$m sources at other wavelengths, 
whilst the
175 $\mu$m survey program has been substantially expanded.\\

Acknowledgements:\\
We would like to thanks the ISOPHOT team for many helpful discussions and
anonymous referee for useful comments.

\end{document}